
\input jytex.tex   
\typesize=10pt
\magnification=1200
\baselineskip=17truept
\hsize=6truein\vsize=8.5truein
\sectionnumstyle{blank}
\chapternumstyle{blank}
\chapternum=1
\sectionnum=1
\pagenum=0

\def\begintitle{\pagenumstyle{blank}\parindent=0pt\begin{narrow}[0.4in]}
\def\endtitle{\end{narrow}\newpage\pagenumstyle{arabic}}


\def\beginexercise{\vskip 20truept\parindent=0pt\begin{narrow}[10
truept]}
\def\endexercise{\vskip 10truept\end{narrow}}


\def\eql#1{\eqno\eqnlabel{#1}}
\def\ref{\reference}
\def\peq{\puteqn}
\def\pref{\putref}

\def\mgn{\marginnote}
\def\bex{\begin{exercise}}
\def\eex{\end{exercise}}

\font\open=msbm10 
\font\opens=msbm8 

\def\mbox#1{{\leavevmode\hbox{#1}}}
\def\hspace#1{{\phantom{\mbox#1}}}
\def\oR{\mbox{\open\char82}}
\def\osR{\mbox{\opens\char82}}
\def\oZ{\mbox{\open\char90}}

\def\rS{{\rm S}}
\def\rHS{{\rm HS}}

\def\al{\alpha}
\def\bbe{{\bmit\beta}} 
\def\bka{{\bmit\kappa}}
\def\be{\beta}

\def\de{\delta}
\def\Ga{\Gamma}

\def\ka{\kappa}
\def\la{\lambda}

\def\om{\omega}

\def\si{\sigma}

\def\th{\theta}
\def\Th{\Theta}
\def\ze{\zeta}

\def\De{\Delta}

\def\Det{{\rm Det\,}}
\def\Real{{\rm Re\,}}
\def\capp{{\rm cap}}
\def\slice{{\rm slice}}
\def\shell{{\rm shell}}
\def\ball{{\rm ball}}

\def\zf{$\zeta$--function}
\def\zfs{$\zeta$--functions}


\def\frac#1/#2{\leavevmode\kern.1em
\raise.5ex\hbox{\the\scriptfont0 #1}\kern-.1em/\kern-.15em
\lower.25ex\hbox{\the\scriptfont0 #2}}
\def\sfrac#1/#2{\leavevmode\kern.1em
\raise.5ex\hbox{\the\scriptscriptfont0 #1}\kern-.1em/\kern-.15em
\lower.25ex\hbox{\the\scriptscriptfont0 #2}}

\def\gtorder{\mathrel{\raise.3ex\hbox{$>$}\mkern-14mu
             \lower0.6ex\hbox{$\sim$}}}
\def\ltorder{\mathrel{\raise.3ex\hbox{$<$}|mkern-14mu
             \lower0.6ex\hbox{\sim$}}}

\def\semidirprod{\rlap{\ss C}\raise1pt\hbox{$\mkern.75mu\times$}}
\def\for{\lower6pt\hbox{$\Big|$}}
\def\fish{\kern-.25em{\phantom{abcde}\over \phantom{abcde}}\kern-.25em}


\def\boxit#1{\vbox{\hrule\hbox{\vrule\kern3pt
        \vbox{\kern3pt#1\kern3pt}\kern3pt\vrule}\hrule}}
\def\dalemb#1#2{{\vbox{\hrule height .#2pt
        \hbox{\vrule width.#2pt height#1pt \kern#1pt
                \vrule width.#2pt}
        \hrule height.#2pt}}}

\def\ds{{|\!|}}        
\def\cd#1{{}_{\ds #1}} 

\def\noin{\noindent}

\def\comb#1#2{{\left(#1\atop#2\right)}}

\def\cosec{{\rm cosec\,}}

\def\eg{{\it e.g. }}
\def\ie{{\it i.e. }}
\def\cf{{\it cf }}
\def\pa{\partial}


\def\tr{{\rm tr\,}}

\def\wR{{\widehat R}}

\def\curl{{\rm curl\,}}

\def\3j#1#2#3#4#5#6{\left\lgroup\matrix{#1&#2&#3\cr#4&#5&#6\cr}
\right\rgroup}

\def\man{{\cal M}}

\def\fd{{\cal F}}

\def\m?{\mgn{?}}


\def\aop#1#2#3{{\it Ann. Phys.} {\bf {#1}} (19{#2}) #3}

\def\cmp#1#2#3{{\it Comm. Math. Phys.} {\bf {#1}} (19{#2}) #3}
\def\cqg#1#2#3{{\it Class. Quant. Grav.} {\bf {#1}} (19{#2}) #3}

\def\jmp#1#2#3{{\it J. Math. Phys.} {\bf {#1}} (19{#2}) #3}
\def\jpa#1#2#3{{\it J. Phys.} {\bf A{#1}} (19{#2}) #3}

\def\np#1#2#3{{\it Nucl. Phys.} {\bf B{#1}} (19{#2}) #3}
\def\pl#1#2#3{{\it Phys. Lett.} {\bf {#1}} (19{#2}) #3}

\def\pr#1#2#3{{\it Phys. Rev.} {\bf {#1}} (19{#2}) #3}
\def\prA#1#2#3{{\it Phys. Rev.} {\bf A{#1}} (19{#2}) #3}

\def\prD#1#2#3{{\it Phys. Rev.} {\bf D{#1}} (19{#2}) #3}

\def\prs#1#2#3{{\it Proc. Roy. Soc.} {\bf A{#1}} (19{#2}) #3}
\def\pcps#1#2#3{{\it Proc. Camb. Phil. Soc.} {\bf{#1}} (19{#2}) #3}

\def\dmj#1#2#3{{\it Duke Math. J.} {\bf {#1}} (19{#2}) #3}

\def\jdg#1#2#3{{\it J. Diff. Geom.} {\bf {#1}} (19{#2}) #3}
\def\jfa#1#2#3{{\it J. Func. Anal.} {\bf {#1}} (19{#2}) #3}

\def\ma#1#2#3{{\it Math. Ann.} {\bf {#1}} ({#2}) #3}

\def\pams#1#2#3{{\it Proc. Am. Math. Soc.} {\bf {#1}} (19{#2}) #3}

\def\tams#1#2#3{{\it Trans. Am. Math. Soc.} {\bf {#1}} (19{#2}) #3}

\begin{title}
\vglue 20truept
\righttext {MUTP/95/1}
\righttext{hep-th/--}
\leftline{\today}
\vskip 100truept
\centertext {\Bigfonts \bf Further functional determinants}
\vskip 15truept
\centertext{J.S.Dowker\footnote{Dowker@a3.ph.man.ac.uk} and J.S.Apps}
\vskip 7truept
\centertext{\it Department of Theoretical Physics,\\
The University of Manchester, Manchester, England.}
\vskip 60truept
\centertext {Abstract}
\begin{narrow}
Functional determinants for the scalar Laplacian on spherical caps and
slices, flat balls, shells and generalised cylinders are evaluated in two,
three and four dimensions using conformal techniques. Both Dirichlet and
Robin boundary conditions are allowed for. Some effects of non-smooth
boundaries are discussed; in particular the 3-hemiball and the
3-hemishell are considered. The edge and vertex contributions to the
$C_{3/2}$ coefficient are examined.
\end{narrow}
\vskip 5truept
\righttext {January 1995}
\vskip 75truept
\righttext{Typeset in \jyTeX}
\vfil
\end{title}
\pagenum=0
\section{\bf 1. Introduction}
This paper is a continuation of our evaluation of the functional
determinant of the (conformal) Laplacian on various regions of the
$d$-sphere and Euclidean
$d$-space. The general idea is to use the behaviour of the functional
determinant under conformal transformations. For the present, this
technique is limited to $d\le4$ because of the unavailability of the
requisite conformal anomaly, $\ze(0)$, -- essentially the $C_{d/2}$
coefficient in the short-time expansion of the heat kernel. The
availability of this coefficient is even further restricted if the
boundary is only piecewise smooth; then, only $C_1$ is known completely.

The object of the present work is to extend the results described in
[\pref{Dow1,Dow2,Dow3}] to higher dimensions and to other regions such
as spherical slices and shells. We also wish to make progress with the
relevant heat-kernel coefficient $(C_{3/2}$) for a {\it piecewise} smooth
boundary. The general form of the coefficients in this case is an open
problem.

The expressions derived for the functional determinants may have some
specific quantum field theoretic application. Caps and balls, for
example, arise in discussions of quantum cosmology  [\pref{Barv,Kam,
Death}]. There are also applications to statistical mechanics in
connection with finite size effects, \eg [\pref{CandC}] and to conformal
field theory, \eg [\pref{GRV}]. In mathematics, the critical points of
the determinant have some
interest, \eg [\pref{Rich,BCY}]. For example, the uniformisation theorem
can be proved using this approach, [\pref{Osgood}].

There are several relevant calculations of functional determinants. We
mention only those by Aurell and Salomonsen on certain simplicial
decompositions, [\pref{Aurell1,Aurell2}], those by Branson and \O rsted,
[\pref{BandO}], and those by Elizalde, [{\pref{Elizalde1}].
\section{\bf 2. Basic equations.}
The functional determinant of the positive elliptic operator $D$ is defined
in terms of the \zf\ of $D$ by the usual relation,
$$
\ln\Det D=-\ze'(0).
$$

For a conformally invariant (scalar) field theory, the conformal behaviour
under infinitesimal Weyl rescalings, $g\to\bar g=\exp(-2\om)g$, is
controlled by the conformal anomaly, $\ze(0)$. Integrated along a
conformal family of metrics, this anomaly yields the finite change in the
functional determinant (or, equivalently, the effective action
$W={1\over2}\ln\Det D$),
$${1\over2}\ln{\Det\overline D\over\Det D}=W[\bar g,g].
\eql{cchange}$$
In two dimensions, the cocycle function $W[\bar g, g]$ has been given by
L\"uscher, Symanzik and Weiss [\pref{Luscher}], Polyakov [\pref{Polyakov}]
(when the boundary is empty) and by Alvarez [\pref{Alvarez}]. The Dirichlet
three-dimensional expression can be found in [\pref{DandS}] (after
correction) as can the four-dimensional one. This last, when there is
no boundary, has been known for some time. Recently, Branson and Gilkey
[\pref{BandG}] have given the four-dimensional result for more general
differential operators and also for Robin boundary conditions (see also
[\pref{Schofield}]). Some related results are given by Dettki and Wipf
[\pref{DandW}].

For completeness, and to correct some errors, the forms of
$W[\bar g,g]$ are given here following [\pref{DandS}].

In three dimensions, and for Dirichlet conditions,
$$\eqalign{
W^D[\bar g,g]={1\over1536\pi}\int_{\pa\man}\bigg[\big(6\tr(\bka^2)
-&3\ka^2-16{\widehat\De}_2\om-4\widehat R\big)\om\cr
&+30\ka N+18N^2-24n^\mu n^\nu \om_{\mu\nu}\bigg],\cr}
\eql{3dd}$$
while in four dimensions
$$\eqalign{
W^D[\bar g,g]=&{1\over2880\pi^2}\int_\man\bigg[(|{\rm Riem}|^2-
|{\rm Ric}|^2+\De_2R)\om-2R_{\mu\nu}\om^\mu\om^\nu\cr
&\hspace{*********}
-4\om^\mu\om_\mu\De_2\om+2(\om^\mu\om_\mu)^2+3(\De_2\om)^2\bigg]\cr
&+{1\over5760\pi^2}\int_{\pa\man}\bigg[\bigg({320\over21}\tr(\bka^3)
-{88\over7}\ka\tr(\bka^2)+{40\over21}\ka^3-4R_{\mu\nu}\chi^{\mu\nu}\cr
&\hspace{*********}
-4\ka R_{\mu\nu}n^\mu n^\nu+16R_{\mu\nu\rho\si}n^\mu n^\rho\chi^{\nu\si}-
2n^\mu\pa_\mu R\bigg)\cr
&\hspace{*********}
-N\bigg({12\over7}\ka^2-{60\over7}\tr(\bka^2)-12\De_2\om+8\om^\mu\om_\mu
\bigg)\cr
&\hspace{*********}-{4\over7}N^2\ka+{16\over21}N^3+24\ka\De_2\om-4\chi^
{\mu\nu}\om_\mu
\om_\nu-\cr
&\hspace{***************}20\ka\om^\mu\om_\mu-30n^\mu\pa_\mu(\De_2\om-\om^
\nu\om_\nu)\bigg]
.\cr}
\eql{4dd}$$

The normal to $\pa\man$, $n^\mu$, points inwards and $\om_\mu=\pa_\mu\om$,
$N=n^\mu\om_\mu$, $\om_{\mu\nu}=\om\cd{\mu\nu}$. The curvature
conventions are those of Hawking and Ellis.
Various equivalent forms can be found upon partial integration or use of
the Gauss-Codazzi equations.

We shall also be concerned with Robin boundary conditions (sometimes
called Neumann by the mathematicians). The corresponding cocycle
function in three dimensions is
$$\eqalign{
W^R[\bar g,g]=-W^D[\bar g,g]+{1\over256\pi}\int_{\pa\man}
\bigg[\big(2\tr&(\bka^2)+\ka^2+32\psi^2-16\psi\ka\big)\om\cr
&+2\big(3\ka-8\psi\big)
n^\mu\om_\mu+2\big(n^\mu\om_\mu\big)^2\bigg].\cr}
\eql{3dr}$$

The four dimensional expression is more lengthy,

$$\eqalign{
W^R[\bar g,g]=&W^D[\bar g,g]\cr
&-{1\over2520\pi^2}\int_{\pa\man}
\bigg[2\big(\tr(\bka^3)-\ka\tr{\bka^2}+{2\over9}\ka^2\big)\om\cr
&+2\,\big(\tr(\bka^2)-{2\over3}\ka^2\big)N-2\ka N^2-2\,N^3\cr
&+{105\over2}\big(4\psi'^3\om-2\psi'^2N
+{1\over3}\psi'(\nabla_2\om-\om^\mu\om_\mu)\big)\cr
&-7(\ka^2-3\tr{\bka^2})(\psi'\om-{1\over6}N)+
7\psi'N(2\ka+3N)\cr
&-{35\over4}\big(3n^\nu\pa_\mu-2\ka\big)(\nabla_2\om-\om^\mu\om_\mu)
\bigg]\cr}
\eql{4dr}$$where $\psi'=\psi-\ka/3$.

An indication of how these equations
are derived is given in section 10. We note that they satisfy the required
symmetry under interchange of $g$ and $\bar g$.

\section{\bf 3. Conformal transformations.}
The conformal transformations under consideration are those between
the \break sphere, $\rS^{d+1}$, Euclidean space, $\oR^{d+1}(\sim\oR^+
\times\rS^d)$ and the cylinder, $\oR\times\rS^d$.

Our general strategy is guided by the fact that it is easier to calculate
the functional determinant on the sphere and the cylinder, than on the
Euclidean ball.

In the present paper we wish to extend some of the results of
[\pref{Dow2}]
to three and four dimensions. This involves the equatorial stereographic
projection, $\rS^{d+1}\to\oR^{d+1}$, expressed by giving the sphere metric
in the conformally-flat form,
$$
d\si^2_{d+1}={4\over(1+r^2)^2}d{\bf r}^2.
\eql{stproj}$$
We shall determine the functional determinant on a 3-ball and a 4-ball,
and,
by an inverse projection, on a spherical cap.
Then we turn to Euclidean spherical shells, which arise on conformally
transforming the generalised cylinder (the `Einstein\break Universe'),
$\oR\times\rS^d$, to Euclidean space. The standard conformally-flat
metric on
the cylinder is
$$
ds^2=d\tau^2+d\si^2_{d}=e^{-2\tau}\big(dr^2+r^2d\si^2_{d}\big)
\eql{cylproj}$$with $r=\exp\tau$.
An inverse stereographic projection would take such a shell to a
{\it slice} of the sphere, $\rS^{d+1}$.

\section{\bf 4. Functional determinants on caps and balls.}

The functional determinant on the $2$-hemisphere
has been determined by Weisberger and the general case is given in
[\pref{Dow3}] (see also [\pref{BandG}] for explicit three- and
four-dimensional expressions). Therefore, for $d=2$ and $d=3$,
(\peq{3dd}) and (\peq{4dd}) may be used to find the functional
determinant on the $3$- and $4$-ball by means of the stereographic
projection, (\peq{stproj}).

It is possible to rescale the ball and project it back onto
the sphere thereby giving a spherical cap, as in [\pref{Dow2}]. A
corresponding application of (\peq{3dd}) yields the cap functional
determinant. The angle of the cap, $\th$, and the radius of the ball,
$a$, are related by
$\th=2\tan^{-1}a$.

We take the opportunity of correcting the result of a transcription error
in [\pref{Dow2}]. Equation (15) of this reference should be replaced by
$$
W^D_{2\rm cap}(\th)-W^D_{2\rm hemisphere}
=-{1\over3}\cos\th-{1\over6}\ln\tan\th/2
\eql{15}$$
with
$$
W^D_{2\rm hemisphere}=-\ze_R'(-1)+{1\over8}-2\ze_R'(0),
$$
and equation (16) by
$$
W^N_{2\rm cap}(\th)-W^N_{2\rm hemisphere}={1\over6}\cos\th+
{1\over12}\ln(1+\cos\th)+{5\over12}\ln(1-\cos\th),
\eql{16}$$ with
$$
W^N_{2\rm hemisphere}=-\ze_R'(-1)+{1\over8}+2\ze_R'(0).
$$

The cocycle functions are
$$
W^N[\bar g,g]={1\over6}(1-\cos\th)-{1\over2}\ln(1+\cos\th)-{1\over3}\ln2
\eql{24}$$
and
$$
W^D[\bar g,g]={1\over6}\ln2-{1\over3}(1-\cos\th).
\eql{14}$$
Figs.1 and 2 contains plots of (\peq{15}) and (\peq{16}) to replace
those in [\pref{Dow2}].

Turning to three dimensions, the unit ball expression is found to be
$$
W^D_{3\ball}=W^D_{3\rm hemisphere}+{7\over64}
\eql{3ball}$$
and for the cap,
$$
W^D_{\rm 3cap}(\th)-W^D_{\rm 3hemisphere}={1\over48}\big(\ln\sin\th+
{21\over4}
\cos^2\th\big).
\eql{3cap}$$

This expression is symmetrical about $\th=\pi/2$ (the hemisphere) where
it has a local minimum. It tends to $-\infty$ at the extremes, $\th=0$
and $\th=\pi$, and has maxima of $\approx0.0745$ when $\sin^2\th=2/21$
($\th\approx 18^\circ, 162^\circ$). A plot is given in Fig.3.

To complete the evaluation of $W_\capp$, the 3-hemisphere effective
action is needed. This has been determined in [\pref{BandG,Dow3}] to
be
$$
W^D_{\rm
3hemisphere}={3\over8}\ze'_R(-2)-{1\over4}\ze_R'(-1)-{1\over16}+
{1\over24}\ln2\approx-0.003682.
\eql{3hemi}$$

The four-dimensional calculation is slightly more involved algebraically
and yields firstly
$$
W^D_{4\ball}=W^D_{4\rm hemisphere}-{1\over180}\ln2-{17\over15120}
\eql{4ball}$$
followed by
$$
W^D_{\rm4cap}(\th)-W^D_{\rm 4hemisphere}=
{1\over180}\bigg({1\over168}\big(1365\cos\th-1399\cos^3\th\big)
+\ln\tan\th/2\bigg).
\eql{4cap}$$
A plot is shown in Fig.4.

To complete the numerics we need the four-hemisphere formula
[\pref{Dow3,BandG}]
$$
W^D_{\rm 4hemisphere}=-{1\over6}\ze'_R(-3)+{1\over4}\ze'_R(-2)
-{1\over12}\ze_R'(-1)-{1\over516}\approx0.003386.
\eql{4hemi}$$
Equation (\peq{4ball}) with (\peq{4hemi}) agrees with corollary 7.2 of
reference [\pref{BandG}].

Some Robin results, calculated using (\peq{3dr}) and (\peq{4dr})
with $\psi=\psi_1$
constant on the surface of the ball, are now presented, firstly for three
dimensions,
$$
W^R_{3\ball}=W^R_{3\rm hemisphere}-{5\over64}+{1-2\psi_1\over8},
\eql{3ballr}$$
$$\eqalign{
W^R_{\rm 3cap}(\th)-&W^R_{3\rm hemisphere}=\cr
&-{1\over48}
\big(1+6(1-2\psi_1)^2\big)\ln\sin\th-{5\over64}\cos^2\th+{1\over8}(1-
2\psi_1)\cos\th.\cr}
\eql{3capr}$$
The value of $\psi$ on the rim of the cap is
$$
\psi_c={1\over\sin\th}\bigg(\big(\psi_1-{1\over2}\big)-{1\over2}\cos\th
\bigg),
\quad {\rm so}\quad 1-2\psi_1=\cos\th-2\psi_c\sin\th,
$$
and that for the hemisphere
$$
\psi_h=\psi_1-{1\over2}.
$$

The geometrical choice is $\psi_1=1/2$, $\psi_c=\cot(\th)/2$ or $\psi_h=0$.
In this case
$$
W^R_{\rm 3cap}(\th)-W^N_{3\rm hemisphere}=
-{1\over48}\ln\sin\th-{5\over64}\cos^2\th
\eql{3caprg}$$
where the {\it Neumann} effective action on the 3-hemisphere is given
in [\pref{Dow3}],
$$
W^N_{3\rm hemisphere}=
{3\over8}\ze_R'(-2)+{1\over4}\ze_R'(-1)+{1\over16}+{1\over12}\ln2
\approx0.067489.
$$

The four dimensional expressions are
$$
W^R_{4\ball}=W^R_{4\rm
hemisphere}+{1\over2160}-{1\over180}\ln2+{1\over12}(\psi-1)^2
+{1\over15}\psi_1-1)\,,
\eql{4ballr}$$
and
$$\eqalign{
W^R_{\rm 4cap}(\th)-&W^R_{4\rm hemisphere}=\cr
&-{1\over180}
\bigg({1\over24}\big(167\cos^3\th-165\cos\th\big)+\ln\tan(\th/2)\bigg)\cr
&+{1\over15}(\psi_1-1)\cos^2\th+{1\over12}(\psi_1-1)^2\cos\th+{1\over6}
(\psi_1-1)^3\ln\sin\th.\cr}
\eql{4capr}$$

The relations between the values of $\psi$ are now,
$$
\psi_1-1=\psi_c\sin\th-\cos\th,\quad{\rm and}\quad
\psi_h=\psi_1-1.
$$
For the geometrical choice, $\psi_c=\cot\th$, \ie $\psi_1=1$, we can use the
expression for the Neumann 4-hemisphere effective action, [\pref{Dow3}],
$$
W^N_{4\rm hemisphere}=
-{1\over3}\ze'_R(-3)-{1\over2}\ze'_R(-2)-{1\over6}\ze'_R(-1)-{1\over288},
$$
if numbers are required.

\section{\bf 5. Symmetry.}
It will be noticed that the $d=3$ result, (\peq{3cap}), is symmetrical
about the hemisphere
($\th=\pi/2$) while those for $d=2$ and $d=4$, (\peq{15}) and
(\peq{4cap}),
are antisymmetrical. It is
possible to show that this is a general feature for odd and even
dimensions. The proof is detailed here. Unless stated otherwise, all
quantities refer to Dirichlet boundary conditions.

The form of the conformal change (\peq{cchange}) is,
in the present instance,
$$W_\capp(\th)-W_\ball(a)=V_0^\th+B_\th
$$where $V$ and $B$ are volume and boundary integrals and we will take
$0\le\th\le\pi/2$. The labels on $V$ are the limits of the colatitude
integration.

For even dimensions,
$$W_\slice(\th,\th')-W_\shell(a,a')=V^{\th'}_\th+B_{\th'}-B_\th
$$ since the normal to the boundary points into the domain, and is thus
oppositely oriented on the two components (the inner and outer faces of the
shell, or slice). For an odd-dimensional boundary, reversing the
orientation reverses the sign of the integral.

Further,
$$W_\ball(a)= W_\ball(1)-\ze_\ball(0)\ln a
$$so we have
$$
W_\capp(\th)+W_\capp(\pi-\th)=2W_\ball(1)+V_0^\th+V_0^{\pi-\th}+B_\th+B_
{\pi-\th}
$$ whence
$$\eqalign{
&\big(W_\capp(\th)-W_{\rm hemisphere}\big)+
\big(W_\capp(\pi-\th)-W_{\rm hemisphere}\big)\cr
&\hspace{**}=\big(V_{\pi/2}^{\pi-\th}+B_{\pi-\th}-B_{\pi/2}\big)-
\big(V_\th^{\pi/2}+B_{\pi/2}-B_\th\big)\cr
&\hspace{**}=\big(W_\slice(\pi/2,\pi-\th)-W_\shell(1,1/a)\big)-
\big(W_\slice(\th,\pi/2)-W_\shell(a,1)\big)\cr
&\hspace{**}=W_\shell(a,1)-W_\shell(1,1/a),\cr}
$$since $W_\slice(\pi/2,\pi-\th)=W_\slice(\th,\pi/2)$ by symmetry.
Hence we
arrive at
$$
\big(W_\capp(\th)-W_{\rm hemisphere}\big)-
\big(W_\capp(\pi-\th)-W_{\rm hemisphere}\big)=-\ze_\shell(0)\ln a=0.
$$
$\ze_\shell(0)$ vanishes because the space is flat, and the boundary terms
cancel, for the same reason as before. Therefore
$W_\capp(\th)-W_{\rm hemisphere}$ is antisymmetrical under
$\th\to\pi-\th$, as stated.

For odd dimensions, there are no volume integrals, and the boundary is even
dimensional. Hence
$$W_\capp(\th)-W_\ball(a)=B_\th$$
and
$$W_\slice(\th,\th')-W_\shell(a,a') = B_\th+B_{\th'}
$$so that
$$\eqalign{
&W_\capp(\th)-W_\capp(\pi-\th)\cr
&\hspace{****}=-2\ze_\ball(0)\ln a+B_\th-B_{\pi-\th}\cr
&\hspace{****}=-2\ze_\ball(0)\ln a+(B_\th+B_{\pi/2})-(B_{\pi/2}
+B_{\pi-\th})\cr
&\hspace{****}=-2\ze_\ball(0)\ln a+\big(W_\slice(\th,\pi/2)
-W_\shell(a,1)\big)\cr
&\hspace{************}-\big(W_\slice(\pi/2,\pi-\th)-W_\shell(1,1/a)\big)\cr
&\hspace{****}=W_\shell(1,1/a)-W_\shell(a,1)-2\ze_\ball(0)\ln a.\cr}
$$ However $\ze_\shell(0)=2\ze_\ball(0)$ since the boundary terms are
equal and there are no volume terms. Therefore
$$\eqalign{
W_\capp(\th)-W_\capp(\pi-\th)&=W_\shell(1,1/a)-W_\shell(a,1)-\ze_\shell(0)
\ln a\cr
&=0\cr}
$$ and $W_\capp(\th)$ is symmetrical under $\th\to\pi-\th$, in odd
dimensions.

We finally note that the three-dimensional Robin result (\peq{3capr}) is
symmetrical about the hemisphere if the values of $\psi$ on the cap
and hemisphere are also reversed under $\th\to \pi-\th$ while the
four-dimensional expression (\peq{4capr}) is antisymmetrical. Again this is a
general feature if $d\ge3$.
\section{\bf 6. Functional determinants on shells and slices}
Weisberger has obtained the effective action on the annulus (or 2-shell)
by
conformal transformation from the cylinder, $I\times\rS^1$.
$$
W^D_{2\shell}=W^D_{I\times\rS^1}-{1\over12}\ln\big(r_1/r_2\big)
$$
where $r_1$ and $r_2$ are the outer and inner radii of the shell.

A conformal
transformation of the $(d+1)$-shell into the $(d+1)$-sphere gives the
effective
action on a $(d+1)$-slice (or spherical, spherical shell). A summary of
our
findings follows.
$$
W^D_{3\shell}=W^D_{I\times\rS^2}+{1\over96}\big(2\ln(r_1 r_2)+9\big),
\eql{3shell}$$

$$
W^D_{4\shell}=W^D_{I\times\rS^3}+{1\over720}\ln\big(r_1/r_2\big),
\eql{4shell}$$

$$
W^D_{2\slice}=W^D_{2\shell}+{2\over3}\sin\Th\sin\De,
\eql{2slice}$$

$$\eqalign{
W^D_{3\slice}=W^D_{3\shell}-&{21\over192}\big(1-\cos2\Th\cos2\De\big)\cr
&+{1\over24}\ln\big(\cos\De+\cos\Th\big),\cr}
\eql{3slice}$$

$$\eqalign{
W^D_{4\slice}=W^D_{4\shell}&-{1\over30240}\sin\Th\sin\De\,\big(2730\cr
&-1399
(2+2\cos2\Th\cos2\De+\cos2\Th+\cos2\De)\big)\cr}
\eql{4slice}$$
where $\Th$ is the colatitude of the midpoint of the slice and $\De$ is
its angular half-width.

We note the geometrical relations
$$\eqalign{
r_1={\sin\Th+\sin\De\over\cos\De+\cos\Th},
&\quad r_2={\sin\Th-\sin\De\over\cos\De+\cos\Th},\cr
r_1r_2={\cos\De-\cos\Th\over\cos\De+\cos\Th},&\quad
L=\ln\big(r_1/r_2\big).\cr}
\eql{georel}$$

As an example, for a slice symmetrical about the equator, $\Th=\pi/2$ and
the cylinder length is $L=\ln\big((1+\sin\De)/(1-\sin\De)\big)$
$=2\tanh^{-1}(\sin\De)$ so that, as $\De\to\pi/2$,
$L\to\infty$ and $\De\to0$ implies $L\to0$.

$W^D_{4\slice}$, (\peq{4slice}), simplifies to
$$
W^D_{4\slice}=W^D_{I\times\rS^3}+{1\over720}\ln
\bigg({1+\sin\De\over1-\sin\De}\bigg)-{1\over15120}\sin\De
\big(1365-1399\sin^2\!\De\big)
$$ and is plotted in Fig.5. It shows a minimum at $\De\approx 46^\circ$.

The general Robin results are more complicated. Corresponding
to (\peq{3shell}),
$$\eqalign{
W^R_{3\shell}=W^R_{I\times\rS^2}-{1\over96}&\big(2\ln(r_1 r_2)+3\big)
+{1\over8}\big((1-2r_1\psi_1)+(1+2r_2\psi_2)\big)\cr
&-{1\over8}
\big((1-2r_1\psi_1)^2\ln r_1+(1+2r_2\psi_2)^2\ln r_2\big),\cr}
\eql{3shellr}$$
where $\psi_1$ and $\psi_2$ are the $\psi$ values on the outer and inner
shell boundaries respectively. For $\psi_1=1/2r_1$ and $\psi_2=-1/2 r_2$,
this simplifies to
$$
W^R_{3\shell}=W^R_{I\times\rS^2}-{1\over96}\big(2\ln(r_1 r_2)+3\big).
$$
The corresponding $\psi$ functions on $I\times \rS^2$ are $r_1\psi_1-1/2$
and
$r_2\psi_2-1/2$, implying Neumann conditions for the geometrical choice.

Transforming the shell to a slice yields,
$$\eqalign{
W^R_{3\slice}&=W^R_{3\shell}+{5\over64}\big(1-\cos2\Th\cos2\De\big)
-{1\over24}\ln\big(\cos\De+\cos\Th\big)\cr
&\hspace{**********}-{1\over8}\big((2\psi_\al\sin\th_1-
\cos\th_1)^2\ln(1+\cos\th_1)\cr
&\hspace{*************}+(2\psi_\be\sin\th_2
+\cos\th_2)^2\ln(1+\cos\th_2)\big)\cr
&+{1\over8}\big((2\psi_\al\sin\th_1-\cos\th_1)(1-\cos\th_1)
-(2\psi_\be\sin\th_2+\cos\th_2)(1-\cos\th_2)\big)\cr}
\eql{3slicer}$$
where $\psi_\al$ and $\psi_\be$ are the $\psi$ values on the slice
boundaries, at the angles $\th_1$ and $\th_2$ respectively.
For $\psi_\al=\cot(\th_1)/2$ and $\psi_\be=\cot(\th_1)/2$ only the first
line
of (\peq{3slicer}) survives and the boundary condition on the interval is
the Neumann one.
$$\eqalign{
W^R_{4\shell}=W^R_{I\times\rS^3}-{1\over6}\big(&(r_1\psi_1-1)^3\ln r_1 +
(r_2\psi_2-1)^3\ln r_2\big)\cr
&+{1\over12}\big((r_1\psi_1-1)^2-(r_2\psi_2-1^2)\big)\cr
&+{1\over40}\big((r_1\psi_1-1)+(r_2\psi_2-1)\big)
+{1\over720}\ln\big(r_1/r_2\big).
\cr}
\eql{4shellr}$$

For the geometrical choice this reduces to the final term which equals
the Dirichlet value.

For the four-dimensional slice, we find
$$\eqalign{
W&^R_{4\slice}-W^R_{4\shell}=\cr
&{1\over4320}\sin\Th\sin\De\,
\big(330-167(2+2\cos2\Th\cos2\De+\cos2\Th+\cos2\De)\big)\cr
&+{1\over6}\bigg(\!\big(\psi_\al\sin\th_1-\cos\th_1\big)^3\ln(1+\cos\th_1)+
\big(\psi_\be\sin\th_2+\cos\th_2\big)^3\ln(1+\cos\th_2)\!\bigg)\cr
&-{1\over12}\bigg(\!\big(\psi_\al\sin\th_1-\cos\th_1\big)^2\ln(1-\cos\th_1)+
\big(\psi_\be\sin\th_2+\cos\th_2\big)^2\ln(1-\cos\th_2)\!\bigg)\cr
&-{1\over15}\bigg(\!\big(\psi_\al\sin\th_1-\cos\th_1\big)\sin^2\th_1+
\big(\psi_\be\sin\th_2+\cos\th_2\big)\sin^2\th_2\!\bigg)
\cr}
\eql{4slicer}$$
with the same notation as before.

\section{\bf 7. Generalised cylinders}
We now find the functional determinant on cylinders
needed to complete the calculation on shells and slices according
to the equations in the previous section. Having some interest in its own
right, it is developed in more generality than is strictly necessary for
present purposes.

The \zf\ on the generalised cylinder, $I\times\man$, is firstly
constructed as
$$\ze^D_{I\times\man}(s)=\sum_{n=1}^\infty\sum_\la{d_\la\over
(\pi^2n^2/L^2+\la^2)^s}
\eql{cylzetad}$$where $L$ is the length of the interval $I$ and $d_\la$
is the degeneracy of
the $\la$ eigenvalue of the scalar Laplacian $-\De_2+\xi R$,
$\xi=(d-1)/4d$, on $\man$, conformal in $(d+1)$-dimensions. For Neumann
conditions, the $n$-summation runs from $0$ upwards, excluding any overall
zero modes.

It is convenient to extend the $n$-summation over negative integers to get
$$
\ze_{I\times\man}(s)={1\over2}\big(\ze_{S^1\times\man}(s)\mp\ze_\man(s)
\big)
$$where, up to a factor, the first \zf\ in the brackets can be thought
of as a thermal one. The signs refer to the boundary conditions
{\it on the interval}.

Standard manipulations similar to those leading to Kronecker's limit formula
(\eg [\pref{Epstein,DandK,Kennedy1,IandZ}]) give
$$
\ze_{I\times\man}'(0)={1\over2}\big(\ze_{\osR\times\man}'(0)\mp
\ze_\man'(0)\big)-\sum_{m=1}^\infty{1\over m}K^{1/2}(2mL)
\eql{fd}$$where
$K^{1/2}(z)$ is the kernel for the pseudo-differential operator
$(-\De_2+\xi R)^{1/2}$. Also
$$
\ze_{\osR\times\man}(s)={1\over\sqrt{4\pi}}{\Ga(s-1/2)
\over\Ga(s)}\ze_\man(s-1/2).
$$

For $\man$ we choose that portion, denoted $\fd$, of $\rS^d$ which acts
as a fundamental domain for the complete symmetry group of a regular
$(d+1)$--polytope classified by the degrees $d_i$, $(i=1,2,\ldots,d_{d+1}
=2)$.

As shown in [\pref{ChandD}], $\ze_\fd(-1/2)$ is finite and, in this case,
$$
\ze_{\osR\times\fd}'(0)=-\ze_\fd(-1/2).
\eql{zedash}$$

The Neumann ($\psi=0$) and Dirichlet  \zfs\ on $\fd$ are explicitly
[\pref{ChandD}],
$$
\ze_\fd(s)=\ze_d\left(2s,a\mid {\bf d}\right),
$$
where the general definition of the Barnes \zf\ is
$$\eqalign{
\zeta_d(s,a\mid{\bf {d}})&={i\Ga(1-s)\over2\pi}\int_L dz {\exp(-a z)
(-z)^{s-1}\over\prod_{i=1}^d\big(1-\exp(-d_i z)\big)}\cr
&=\sum_{{\bf {m}}={\bf 0}}^\infty{1\over(a+{\bf {m.d}})^s},\qquad
\Real\, s>d,\cr}
\eql{barnes}$$
which shows that the eigenvalues are perfect squares,
$$
\la_n=(a+{\bf m.d})^2,
\eql{eigenv}$$the degeneracies arising from coincidences. It is then easy
to
construct the kernel $K^{1/2}$. It is evident in (\peq{barnes}).
The parameter $a$ is $(d-1)/2$ for Neumann conditions and
$\sum d_i-(d-1)/2$ for Dirichlet. There are no zero modes if $d>1$.

Equation (\peq{fd}) becomes
$$
\ze_{I\times\fd}'(0)=-{1\over2}\ze_d\big(-1,a\mid {\bf d}\big)
\mp\ze_d'\big(0,a\mid {\bf d}\big)-
\sum_{m=1}^\infty{e^{-2amL}\over m}\prod_{i=1}^d{1\over1-q_i^m}
\eql{fd2}$$where $q_i=\exp(-2Ld_i)$.

Barnes has given formulae for $\ze_d(-n)$, $n\in\oZ$, in terms of
generalized Bernoulli functions. In particular
$$\ze_d\big(-1,a\mid {\bf
d}\big)={(-1)^d\over\prod_id_i}{1\over(d-1)!}B^{(d)}_{d+1}
\big(a\mid{\bf d}\big)
$$
which is needed in (\peq{zedash}).

The derivative of the Barnes \zf\ is related to the multiple gamma
function, [\pref{Barnesa}]. A method of evaluation is contained in
[\pref{Dow1}] and so we may assume that the functional determinant on the
generalised cylinder, $I\times\fd$, is known. Unfortunately, as mentioned
earlier, because the heat-kernel coefficients are not
extant in the piecewise smooth case if $d\ge3$, the functional determinant
cannot be conformally transformed unless $\fd$ is the
complete sphere. Of course, in this case, it is not necessary to invoke
the full generality of the Barnes \zf. The formulae could be derived
directly and have been known for a long time.

The hemisphere degrees are all unity, ${\bf d=1}$, and the \zf\ reduces to

$$\eqalign{
\ze_d(s,a)&={i\Ga(1-s)\over2\pi}\int_L{e^{z(d/2-a)}(-z)^{s-1}\over
2^d\sinh^d(z/2)}\,dz\cr
&=\sum_{m=0}^\infty\comb{m+d-1}{d-1}{1\over{(a+m)}^s}\,,\quad \Real s>d,\cr
}
\eql{hemibarnes}$$ with $a=(d+1)/2$ for Dirichlet (D) and $a=(d-1)/2$ for
Neumann (N) conditions {\it on the hemisphere rim}.

The summation form can be adjusted to the following expressions,
$$\eqalign{
\ze^D_d(s)&={1\over(d-1)!}\sum_{m=1}^\infty{(m+q-1)\ldots(m-q)\over m^s}
,\cr
\ze^N_d(s)&={1\over(d-1)!}\sum_{m=1}^\infty{(m+q)\ldots(m-q+1)\over m^s}
,\cr}
\eql{hemizetao}$$for odd dimensions ($d=2q+1$), while for even dimensions
($d=2q+2$)
$$\eqalign{
\ze^D_d(s)&={1\over(d-1)!}\sum_{m=0}^\infty{(m+q)\ldots(m-q)
\over(m+1/2)^s},\cr
\ze^N_d(s)&={1\over(d-1)!}\sum_{m=0}^\infty{(m+q+1)\ldots(m-q+1)
\over(m+1/2)^s}.\cr}
\eql{hemizetae}$$

We therefore need the Stirling number expansions
$$(m+a)(m+a-1)\ldots(m+a-b+1)=\sum_{k=0}^bS^{(k)}(a,b)\,m^k=
\sum_{k=0}^bT^{(k)}(a,b)\,(m+1/2)^k$$
which allow the series in (\peq{hemizetao}), (\peq{hemizetae}) to be
written as sums of
Riemann \zfs\ so providing a practical continuation in a familiar fashion.

For odd $d$,
$$\eqalign{
\ze^D_d(s)&=\sum_{k=0}^{2q}S^{(k)}(q-1,2q)\,\ze_R(s-k),\cr
\ze^N_d(s)&=\sum_{k=0}^{2q}S^{(k)}(q,2q)\,\ze_R(s-k),\cr}
$$and for even $d$,
$$\eqalign{
\ze^D_d(s)&=\sum_{k=0}^{2q+1}T^{(k)}(q,2q+1)\,\ze_R(s-k,1/2),\cr
\ze^N_d(s)&=\sum_{k=0}^{2q+1}T^{(k)}(q+1,2q+1)\,\ze_R(s-k,1/2).\cr}
\eql{hemizeta2}$$

A few expressions are made explicit:
$$\eqalign{
&\ze_2(s)=\ze_R(s-1,1/2)\mp{1\over2}\ze_R(s,1/2),\cr
&\ze_3(s)={1\over2}\big(\ze_R(s-2)\mp\ze_R(s-1)\big),\cr
&\ze_4(s)={1\over6}\big(\ze(s-3,1/2)\mp{3\over2}\ze_R(s-2,1/2)
-{1\over4}\ze_R(s-1,1/2)\pm{3\over8}\ze_R(s,1/2)\big),\cr
&\ze_5(s)={1\over24}\big(\ze(s-4)\mp2\ze_R(s-3)
-\ze_R(s-2)\pm2\ze_R(s-1)\big),\cr
&\ze_6(s)={1\over5!}\big(\ze_R(s-5,1/2)\mp{5\over2}\ze_R(s-4,1/2)
-{5\over2}\ze_R(s-3,1/2)\cr
&\hspace{***********}\pm{25\over4}\ze_R(s-2,1/2)+
{9\over16}\ze_R(s-1,1/2)\mp{45\over12}\ze_R(s,1/2)\big),\cr
&\ze_7(s)={1\over6!}\big(\ze_R(s-6)\mp3\ze_R(s-5)-5\ze_R(s-4)\cr
&\hspace{***************}\pm15\ze_R(s-3)+4\ze_R(s-2)\mp
12\ze_R(s-1)\big),\cr}
\eql{hemizeta3}$$where the upper sign refers to Dirichlet conditions on
the hemisphere rim and the lower one to Neumann. The {\it full} sphere
expression is found simply by adding these two forms.

It is now straightforward to determine $\ze'_d(0)$ and $\ze_d(-1)$ in terms
of named zeta functions and the cylinder functional determinant then
follows from (\peq{fd2}), the final term being readily evaluated.

To complete the conformal derivation of the functional
determinants on shells and slices, more explicit values will be exhibited
in the two cases of numerical relevance here:
$$
\ze'_{I\times\rHS^2}(0)=
{1\over2}\ze_R'(-1)+\bigg({1\over24}(1\mp6)\ln2\mp{1\over96}\bigg)
-{1\over4}\sum_{m=1}^\infty{e^{\pm mL}\over m\sinh^2\!mL}
\eql{zedash2}$$
and
$$
\ze'_{I\times\rHS^3}(0)=-\bigg({1\over2}\ze_R'(-2)\mp{1\over2}
\ze_R'(-1)\bigg)-{1\over480}
-{1\over8}\sum_{m=1}^\infty{e^{\pm mL}\over m\sinh^3\!mL}
\eql{zedash3}$$
with $q=\exp(-2L)$. The above expressions are for Dirichlet conditions on
$I$. For Neumann conditions the terms in large brackets are to be reversed
in sign.

Therefore, \cf [\pref{ChandD}] for the final term,
$$\eqalign{
&W^{D,N}_{I\times\rS^2}=\mp{1\over2}\ze'_R(-1)\mp{1\over24}\ln2+{1\over4}
\sum_{m=1}^\infty{\cosh mL\over m\sinh^2\!mL}\cr
&W^{D,N}_{I\times\rS^3}=\pm{1\over2}\ze'_R(-2)+{1\over480}+{1\over8}
\sum_{m=1}^\infty{\cosh mL\over m\sinh^3\!mL}\cr}
\eql{cylea}$$
which are required in (\peq{3shell}) and (\peq{4shell}). Now, of course,
$N$ and $D$ refer to interval boundary conditions. For large $L$
the summations vanish exponentially while for small $L$ one gets the
typical Planckian high temperature forms.

\section{\bf 8. Non-smooth boundaries}
Although $C_{3/2}$ is not known for non-smooth boundaries,
it is possible to make a certain amount of progress since
dimensions restricts the unknown contributions. For Dirichlet conditions,
in $d$-dimensions, we conjecture that
$$\eqalign{
C^D_{3/2}=&{\sqrt\pi\over192}\sum_i\int_{\pa{\cal M}_i}
\big(6\tr(\bka_i^2)-3\ka_i^2-4\wR+12(8\xi-1)R\big)\cr
&-{\sqrt\pi\over24}\sum_{(ij)}\int_{E_{ij}}\bigg[
\la(\th_{ij})\,(\ka_i+\ka_j)+\mu(\th_{ij})\,(\ka^{(i)}+\ka^{(j)})\bigg]
+\sum_l\int_{V_l}\nu\cr}
\eql{3/2}$$where
$\ka^{(i)}$ and $\ka^{(j)}$ are the traces of two extrinsic curvatures,
$\bka^{(i)}$ and $\bka^{(j)}$, associated with the codimension-2
intersection,
$E_{ij}=\pa{\cal M}_i\cap\pa{\cal M}_j$.
$\bka^{(i)}$ can also be interpreted as the
extrinsic curvature of $\pa{\cal M}_i\cap\pa{\cal M}_j$ considered as a
codimension-1 submanifold of $\pa\man_i$ -- part of the
`boundary of a boundary'. $\ka_i$ and $\ka_j$ are the traces of the
extrinsic curvatures, $\bka_i$ and $\bka_j$, of the boundary parts,
$\pa\man_i$
and $\pa\man_j$ respectively. $\th_{ij}$ is the dihedral angle between
these
two parts. (It can vary along the intersection.) The integrand, $\nu$, of
the
vertex (or `corner') contribution is a function of the dihedral angles
between those boundary parts meeting at the vertex.

It is a theorem that, for a flat ambient space, the extrinsic
curvatures and the normal fundamental forms define an embedded submanifold
up to Euclidean motions if the Codazzi-Mainardi and Ricci conditions are
satisfied, \eg [\pref{Spivak}] IV p 64. Thus, in general, we might expect
the heat-kernel coefficients to depend on the
normal fundamental 1-form, $\bbe$, of $E$ considered as a codimension-2
submanifold of $\man$. This is ruled out immediately on dimensional
grounds since it would have to occur in $C_{3/2}$ in the gauge invariant
combination $\curl\bbe{\bf.}\curl\bbe$. We note that $\bbe$ is conformally
invariant.

Some restrictions follow from the requirement that $C^D_{3/2}$ be
conformally invariant, for $\xi=\xi(d)=(d-2)/4(d-1)$, when $d=3$.
Applying the standard conformal transfomations, we find the relation
$$
2\tan(\th/2)\la(\th)+\mu(\th)=1.
\eql{reln}$$

In two dimensions the $\ka^{(i)}$ are zero and the result for the disc
shows that $\la(\pi)=0$, hence $\mu(\pi)$ is finite.
If $\th_{ij}=\pi$, $\ka^{(i)}+\ka^{(j)}=0$.
(When $\th_{ij}=\pi$ the join between $\pa\man_i$ and
$\pa\man_j$ is not necessarily smooth.)

To prove (\peq{reln}), the required conformal transformations are
$$
(2\tr\bka_i^2-\ka_i^2)\to e^{2\om}\big(2\tr\bka_i^2-\ka_i^2-2(d-3)\ka_i N_i
-(d-3)(d-1)N_i^2\big)
\eql{kachange}$$

$$
\widehat R\to e^{2\om}\big(\widehat
R+2(d-2)\widehat\De_2\om-(d-2)(d-3)\om^i\om_i\big)
\eql{Rchange}$$
where $N_i=n^\mu_i\om_\mu$ and $n^\mu_i$ is the inward normal to
$\pa\man_i$.

We are looking for the change in $C_{3/2}$ to vanish when $d=3$ hence
anything with a factor of $(d-3)$ can be ignored (for this calculation).
Note $8\xi-1=(d-3)/(d-1)$. Therefore in the $\pa\man$ part, only the
$\widehat\De_2\om$ term from (\peq{Rchange}) remains.
This is
$$-{\sqrt\pi\over24}(d-2)\int_{\pa\man_i}e^{(3-d)\om}\widehat\De_2\om.
$$

Now integrate by parts once,
$$
-{\sqrt\pi\over24}(d-2)(3-d)\int_{\pa\man_i}e^{(3-d)\om}h^{ab}\pa_a
\om\pa_b\om
+{\sqrt\pi\over24}(d-2)\sum_j\int_{E_{ij}}e^{(3-d)\om}N_{(i)}
\eql{parint}$$
where $N_{(i)}=n^\mu_{(i)}\om_\mu$ and $n^\mu_{(i)}$ is the normal
(to the edge $E_{ij}$) that points into the face $\pa\man_i$.
(Any necessary summations over $i$ have been dropped,
and $h^{ab}$ is the intrinsic metric on $\pa\man$.)

The last term in (\peq{parint}) is not conformally invariant at $d=3$. We
expect it to be cancelled by the changes in the edge integrals. For
these, the new conformal transformations needed are
$$
\ka^{(i)}\to e^{\om}\big(\ka^{(i)}+(d-2)N_{(i)}\big)
\eql{kachange2}$$and
$$
d{\rm vol}_I\to e^{(2-d)\om}d{\rm vol}_I.
$$
Also
$$
\ka_i\to e^{\om}\big(\ka_i+(d-1)\,n^\mu_i\om_\mu\big).
$$

In terms of the dihedral angle $\th$ there is the geometrical
relation (all the $n$'s are
normalised),
$$ n^\mu_i\big|_E=\cosec(\th_{ij})\, n^\mu_{(j)}-\cot(\th_{ij})\,
n^\mu_{(i)}
\eql{normrel1}$$
so that
$$
\big(n^\mu_i+n^\mu_j\big)\big|_E=\tan(\th_{ij}/2)\,(n^\mu_{(i)}+n^\mu_
{(j)}).
\eql{normrel2}$$
Requiring the nonconformally invariant terms to cancel at $d=3$ yields
(\peq{reln}).

Information about $\la$, $\mu$ and $\nu$ follows in traditional fashion
by special
case evaluation. In two dimensions, the $\bka^{(i)}$ are zero. The
heat-kernel
expansion can be derived without difficulty for a sector of a disc, in
particular for half a disc, $HD^2$. From this we find that
$$
\la(\pi/2)=-3.
\eql{laval}$$

The expansion for the cylinder, $I\times D^2$, follows by trivial
product and yields
$$\mu(\pi/2)=7
\eql{muval}$$ agreeing with (\peq{reln}).

These values are sufficient to determine the effective action when there
are no corners and the contiguous boundary parts are perpendicular. Such
will be the case for the cylinder, $I\times\,{\rm hemisphere}$. The
results of this evaluation are given in section 9.
\subsection{\it Corner contributions}
The expansion on the polygonal cylinder, $I\times{\rm polygon}$, is easily
deduced from that on the polygon and yields the particular values for the
constant term in the heat-kernel expansion, [\pref{Pathria,Baltes}],
$$
w\big({\pi\over\th},2,2\big)=\mp{1\over96}\bigg({\pi\over\th}-
{\th\over\pi}\bigg),
$$
where the notation is $w\big(\pi/\th_1,\pi/\th_2,\pi/\th_3\big)$ in
terms of the three dihedral angles $\th_1,\th_2$ and $\th_3$.

Other specific values for trihedral corner contributions in three
dimensions have been evaluated in [\pref{Dow5}]. The constant term is
$$
w(3,3,2)=\mp1/16,\quad w(3,4,2)=\mp15/128,\quad w(3,5,2)=\mp15/64.
$$
\subsection{\it The smeared coefficient}
Conformal variation can be employed to determine the useful smeared
coefficient
$$
C^{(d)}_k[g;f]\equiv\int_\man C^{(d)}_k(g,x,x)f(x)
$$ by the formula (cf [\pref{DandS}])
$$
C^{(d)}_k\big[g;\de\om\big]=-{1\over d-k/2}\de C^{(d)}_k
\big[e^{-2\om}g;1\big]\big|_{\om=0}-2C^{(d)}_{k-1}(g;{\bf J}\de\om)
\eql{conftr}$$
derived from the variation of the zeta function, \eg [\pref{Dow7}].
${\bf J}$ is the operator
$$
{\bf J}=(d-1)\big(\xi-\xi(d)\big)\De_2.
$$

A straightforward calculation yields
$$\eqalign{
C^{(d)}_{3/2}[g;f]=&{\sqrt\pi\over192}\sum_i\int_{\pa{\cal M}_i}\bigg[
\big(6\tr(\bka_i^2)-3\ka_i^2-4\wR+12(8\xi-1)R\big)f\cr
&\hspace{**********}
+30\ka n^\mu_if_\mu-24n^\mu_i n^\nu_i f_{\mu\nu}\bigg]\cr
&-{\sqrt\pi\over24}\sum_{(ij)}\int_{E_{ij}}\!\bigg[
\la(\th_{ij})\,(\ka_i+\ka_j)f+\mu(\th_{ij})\,(\ka^{(i)}+\ka^{(j)})f\cr
&\hspace{***********}-{1\over2}\big(\mu(\th_{ij})+5\big)\,
(n^\mu_{(i)}+n^\mu_{(j)})f_\mu\bigg]+\sum_l\int_{V_l}\nu\,f,\cr}
\eql{sm3/2}$$
where $f_\mu=\pa_\mu f$ and $f_{\mu\nu}=f\cd{\mu\nu}$.
The first line (\ie the smooth expression) can be found in [\pref{BandG2,
DandS}].

A small technical point is that the second part of the right-hand side of
(\peq{conftr}) removes the $\xi$ dependence from the $\De_2\om$ term that
results from the variation of the $R$ term in the first part, and replaces
it with $\xi(d)$, confirming that the edge terms in (\peq{sm3/2}) do not
depend on $\xi$.
\section{\bf 9. Robin boundary conditions}
The expression for the Robin $C_{3/2}$ is given in [\pref{KCD}], eqn.
5.11c, in the smooth case as
$$
C^R_{3/2}=-C^D_{3/2}+
{\sqrt\pi\over32}\int_{\pa{\cal M}}
\big(2\tr(\bka^2)+\ka^2+32\psi^2-16\psi\ka\big)
\eql{R3/2}$$
where $\psi$ is the function in the
boundary condition $(n^\mu\pa_\mu\phi-\psi\phi)\big|_{\pa\man}=0$
obeying the compensating conformal transformation
$$
\psi\to e^\om\big(\psi+{1\over2}(d-2)\,n^\mu\om_\mu\big).
\eql{psiconf}$$
It is readily checked that the second term in (\peq{R3/2}) is conformally
invariant at $d=3$.

A purely geometrical choice for $\psi$ is, in each dimension
[\pref{KCD, Kennedy2}]
$$
\psi={(d-2)\over2(d-1)}\,\ka
\eql{hkbc1}$$
and in this case the integrand in $(\peq{R3/2}$) becomes
$$12\tr(\bka^2)-6\big(1-2(d-3)^2\big)\ka^2$$
evaluating to the conformally invariant combination, $12\tr(\bka^2)-6
\ka^2$, at
$d=3$.

For a non-smooth boundary, the Robin condition applied to each piece
is
$$
\big(n^\mu_i\pa_\mu\phi-\psi_i\phi\big)\big|_{\pa\man_i}=0.
\eql{neum1}$$

Applying the condition directly to the intersection $E_{ij}$,
$$
\big(n^\mu_{(i)}\pa_\mu\phi-\psi^{(i)}\phi\big)\big|_{E_{ij}}=0
\eql{neum2}$$
and the general relation,
$$
(\psi_i+\psi_j)\big|_I=\tan(\th_{ij}/2)\big(\psi^{(i)}+\psi^{(j)}\big),
\eql{psirel}$$
follows from (\peq{normrel2}).

The geometrical choice (\peq{hkbc1}) on each boundary piece reads,
$$
\psi_i={(d-2)\over2(d-1)}\,\ka_i\,.
\eql{hkbc2}$$
On $E_{ij}$ there is another choice. We could set, referring to
(\peq{kachange2}),
$$
\psi^{(i)}={1\over2}\ka^{(i)}.
\eql{hkbc3}$$

Relation (\peq{psirel}) is not consistent with both
(\peq{hkbc2}) {\it and} (\peq{hkbc3}) being satisfied.
Presumably (\peq{hkbc2}) is the the physically relevant choice, inducing
the
boundary condition on $I$ via
$$
\psi^{(i)}=\big(\cosec(\th_{ij})\,\psi_j + \cot(\th_{ij})\,\psi_i\big)
\big|_E\,.
$$

Given only one (arbitrary smooth) submanifold, then of course only
intrinsically defined objects are available. It is an interesting problem
to extend the usual theory of heat-kernel expansions to the case when the
field satisfies some condition on a submanifold of {\it any} codimension.

If the geometric choice (\peq{hkbc2}) is not made, then it would seem
simplest to assume that $\psi$ is continuous over all of $\pa\man$ so
that $\psi^{(i)}=\psi^{(j)}$.

Taking the Dirichlet expression as exhibited
in (\peq{3/2}), there will be an additional edge term involving $\psi$,
$$\eqalign{
C^R_{3/2}=-C^D_{3/2}+
{\sqrt\pi\over192}\sum_i\int_{\pa{\cal M}_i}\big(12\tr(\bka_i^2)
&+6\ka_i^2+192\psi_i^2-96\psi_i\ka_i\big)\cr
&+{\sqrt\pi\over12}\sum_{(ij)}\int_{E_{ij}}\rho(\th_{ij})
\big(\psi^{(i)}+\psi^{(j)}\big)\,.\cr}
\eql{R3/2ns}$$
Relation (\peq{psirel}) shows that the final term is general enough.

This time, conformal invariance at $d=3$ gives the relation
$$
2\tan(\th/2)\la(\th)+\mu(\th)+\rho(\th)=1.
\eql{rreln}$$

Moss [\pref{Moss}] has considered the Robin case (with $\psi$ constant
which is adequate here) on the disc and hemisphere. The expressions show
immediately that $\rho(\pi)=0$. Extending his results to the
two-dimensional hemidisc and the 3-cylinder reveals that
$$
\la(\pi/2)=9,\quad\mu(\pi/2)=-5,\quad\rho(\pi/2)=-12,
\eql{rvals}$$ with checks.

The smeared coefficient follows after a conformal transformation,
$$\eqalign{
C^{R}_{3/2}[g;f]=-C^{D}_{3/2}[g;f]+&{\sqrt\pi\over32}
\sum_i\int_{\pa\man_i}\bigg(\big(2\tr(\bka_i^2)
+\ka_i^2+32\psi_i^2-16\psi_i\ka_i\big)f\cr
&\hspace{*****************}+12(8\psi_i-3\ka_i)n^\mu_{(i)}f_\mu\bigg)\cr
&+{\sqrt\pi\over24}\sum_{(ij)}\int_{E_{ij}}\!\rho(\th_{ij})\bigg(2
\big(\psi^{(i)}+\psi^{(j)}\big)f+(n^\mu_{(i)}+n^\mu_{(j)})f_\mu\bigg)\cr}
\eql{sm3/2r}$$ where $C^{D}_{3/2}[g;f]$ is given by (\peq{sm3/2}).
\section{\bf 10. The cocycle function}
In the conformally invariant case, there are two equivalent ways of
deriving the cocycle function, $W[\bar g,g]$. One consists of integrating
the conformal anomaly equation,
$$
\de W[\bar g]={1\over(4\pi)^{3/2}}\,C^{(3)}_{3/2}\big[\bar g;\de\om\big]
\eql{can}$$
and the other involves a finite transformation in general dimensions,
the formula being
$$
W[e^{-2\om}g,g]=\lim_{d\to
3}(4\pi)^{-d/2}\,{C^{(d)}_{3/2}[e^{-2\om}g]-C^{(d)}_{3/2}[g]\over d-3}.
\eql{effct}$$

For smooth boundaries, equation (\peq{3dd}) gives the result, which we
will now refer to as $W^D_S[\bar g,g]$. The additional effects of edges
and corners is contained in
$$\eqalign{
W^D[\bar g,g]=W^D_S&[\bar g,g]-{1\over384\pi}\sum_{(ij)}\int_{E_{ij}}
\bigg[2\big(\la(\ka_i+\ka_j)+\mu(\ka^{(i)}+\ka^{(j)})\big)\om\bigg.\cr
&\bigg.-(5+\mu)\,(n^\mu_{(i)}+n^\mu_{(j)})\,\om_\mu
+4\om\big(n^\mu_{(i)}+n^\mu_{(j)}\big)
\om_\mu\bigg]+\sum_l\int_{V_l}\nu\,\om.\cr}
\eql{cfactor}$$

The Robin cocyle function is determined to be
$$\eqalign{
W^R[\bar g,g]=-W^D&[\bar g,g]
+{1\over256\pi}\sum_i\int_{\pa\man_i}\bigg[
\big(2\tr(\bka_i^2)+\ka_i^2+32\psi_i^2
-16\psi_i\ka_i\big)\om\cr
&\hspace{*************}+2(3\ka-8\psi) n^\mu_i\om_\mu+2(n^\mu_i\om_\mu)^2
\bigg]\cr
&+{1\over384\pi}\sum_{(ij)}\int_{E_{ij}}
\bigg[2\rho\,(\psi^{(i)}+\psi^{(j)})\om
-\rho\,(n^\mu_{(i)}+n^\mu_{(j)})\,\om_\mu\bigg]\cr}
\eql{rcfactor}$$
where $W^D$ is given by (\peq{cfactor})

These formulae, together with the values (\peq{laval}), (\peq{muval}) and
(\peq{rvals}) can be used to find the effective action on a 3-hemiball
from that on a quarter 3-sphere and also on a 3-hemishell from that on
the cylinder $I\times$ 3-hemisphere. We find
$$\eqalign{
&W^D_{3\rm hemiball}=W^D_{{1\over4}-3\rm sphere}+
{1\over384}(53-4\ln2)+{1\over48}\ln a\cr
&W^D_{3\rm hemicap}=W^D_{{1\over4}-3\rm sphere}+
{1\over96}\big(\ln(1-\cos\th)+8\cos\th+{21\over4}\cos^2\th\big)\cr
&W^D_{3\rm hemishell}=W^D_{I\times2\rm hemisphere}+
{1\over96}\ln\big({r_1^3\over r_2}\big)+{9\over192}\cr
&W^D_{3\rm hemislice}=W^D_{3\rm hemishell}+
{1\over96}\bigg[2\ln(\cos\Th+\cos\De)-16\sin\Th\sin\De\cr
&\hspace{***********************} +{21\over4}
(\cos2\Th\cos2\De-1)\bigg]\cr}
\eql{3hemis}$$
with the same geometrical relations as before, (\peq{georel}).
For space reasons, the corresponding Robin expressions are not given.

Since $W_{I\times2\rm hemisphere}$ is contained in (\peq{zedash2}), the
only unknown quantity on the right-hand side of (\peq{3hemis}) is
$W_{{1\over4}-3\rm sphere}$. This can be obtained by the methods of
[\pref{Dow1,Dow3}].

\section{\bf 10. The quartersphere effective action}

In this case, there are two perpendicular reflecting hyperplanes and all
the degrees are unity, except for $d_1=2$. The \zf\ reduces to
$$
\ze_{QS}(s,a)=
\sum_{m,n=0}^\infty\comb{m+d-2}{d-2}{1\over{(a+2n+m)}^{2s}-\al^2}\,,
\quad \Real s>d,
\eql{quarterzeta}$$ where $\al=1/2$ for conformal coupling in
$d$-dimensions. This time $a=(d+3)/2$ for Dirichlet (D) and
$a=(d-1)/2$ for Neumann (N) conditions. Again, for brevity, only the
Dirichlet forms are exposed.

The techniques of [\pref{Dow1,Dow3}], applied to (\peq{quarterzeta}) yield
expressions involving the Barnes \zf. One finds for the relevant
quantity, (\cf eqn.(23) of [\pref{Dow3}]),
$$
\ze_{QS}'(0)=\ze_d'(0,d/2+1)+\ze_d'(0,d/2+2)-\sum_{r=1}^u{1\over2^{2r}r}
N_{2r}(d)\sum_{k=0}^{r-1}{1\over 2k+1}.
\eql{qzetadash0}$$
Here $u$ equals $d/2$ if $d$ is even, and $(d-1)/2$ if $d$ is odd.
$N$ is the residue of the Barnes \zf,
$$
\ze_d(s+r,a)\rightarrow{N_r(d)\over s}\quad{\rm as}\,\,s\rightarrow0,
\eql{bresidue}$$where, in this case,
$$\eqalign{
\ze_d(s,a)&=
{i\Ga(1-s)\over2\pi}\int_L{e^{z(d/2+1/2-a)}(-z)^{s-1}\over
2^d\sinh^{d-1}(z/2)\sinh(z)}\,dz.\cr
&=\sum_{m,n=0}^\infty\comb{m+d-2}{d-2}{1\over{(a+2n+m)}^s}\,,
\quad \Real s>d.\cr}
\eql{quarterbarnes}$$

$N$ depends on $a$ and is given by a generalized Bernoulli polynomial.
In the present case it is easiest to find the residue directly from the
integral form of $\ze_d$.

For $d=3$ a standard evaluation gives
$$
\sum_{r=1}^u{1\over2^{2r}r}N_{2r}(d)\sum_{k=0}^{r-1}{1\over 2k+1}=
{1\over4}N_{2}(3)=-{1\over8}.
\eql{resi3}$$

To complete (\peq{qzetadash0}), it is necessary to calculate the first two
terms on the right-hand side. Noting that $a=d/2+1$ or $a=d/2+2$, it is
convenient to adjust the summation over $m$ in (\peq{quarterbarnes}) to
allow for the extra 1 or 2 and to keep the lower limit equal to 0. Then
$$\eqalign{
\ze_d\big(s,{d\over2}+1\big)+\ze_d\big(s,{d\over2}+2\big)&=2S_d(s)
-\sum_{n=0}^\infty\bigg[{2\over(d/2+2n)^s}+
{d-1\over(d/2+1+2n)^s}\bigg]\cr
&=2S_d(s)-2^{-s}\bigg(2\ze_R\big(s,{d\over4}\big)
+(d-1)\ze_R\big(s,{d\over4}+{1\over2}\big)\!\bigg)\cr}
\eql{qbarns}$$
where, after rearrangement, the summation, $S_d(s)$, is
$$\eqalign{
S_d(s)&=
\sum_{m,n=0}^\infty\comb{m+d-2}{d-2}{1\over{(d/2+2n+m)}^s}\cr
&={1\over(d-2)!}\sum_{n=0,m=1}^\infty{(m+q-1)\ldots(m-q)
\over \big(2n+m\big)^s},\cr}
\eql{quzetae}$$for even dimensions ($d=2q+2$), while for odd dimensions,
($d=2q+3$),
$$
S_d(s)={1\over(d-2)!}\sum_{n,m=0}^\infty{(m+q)\ldots(m-q)
\over(2n+m+1/2)^s}.
\eql{quzetao}$$

To reduce the summations further, we introduce, as in [\pref{Dow1}],
the residue classes $m=2l+p$ with $0\le l\le\infty$ and $p=0,1$, treating
the different $p$ values separately ($m$ even and $m$ odd). The
denominator functions read $2(l+n)+$ constant, so we
set $N=l+n$ and effect the sum over $\nu=l-n$ algebraically using
$$
\sum_{n,m=0}=\sum_{N=0}^\infty\sum_{\nu=-N}^N.
$$

For arbitrary dimensions, some general rearrangement of the numerators
is needed so, for rapidity, attention is restricted to $d=3$, \ie $q=0$,
when the numerator in (\peq{quzetao}) is simply
$m^2=\nu^2+2\nu(N+p)+2Np+p^2+N^2$. The term odd in $\nu$ will sum to zero
leaving,
$$\eqalign{
S_3(s)&=\sum_{N=0,p=0,1}^\infty{(2N+1)(p^2+2Np+N^2+N(N+1)/3)
\over(2N+p+1/2)^s}\cr
&={2^{-s}\over12}\big(4\ze_R(s-2,{1\over4})+4\ze_R(s-2,{3\over4})
-\ze_R(s-1,{1\over4})-\ze_R(s-1,{3\over4})\big)\cr
&={2^{-s}\over12}\bigg[4Z\big(s-2,{1\over4}\big)-Z\big(s-1,{1\over4}\big)
\bigg],\cr}
$$
where
$$
Z(s,x)=\sum_{m=-\infty}^\infty{1\over|x+m|^s}=\ze_R(s,x)+\ze_R(s,1-x)
$$
is an Epstein \zf, which is usually more convenient for manipulation. We
have the relation
$$
2^{-s}Z(s,1/4)=\ze_R(s,1/2)=(2^s-1)\ze_R(s).
$$

At $d=3$ the right-hand side of (\peq{qbarns}) becomes, finally,
$$\eqalign{
&{2^{-s}\over6}\bigg[4Z\big(s-2,{1\over4}\big)-Z\big(s-1,{1\over4}\big)
-12Z\big(s,{1\over4}\big)\bigg]+2^{1+s}\cr
&={1\over6}\bigg[4\ze_R\big(s-2,{1\over2}\big)-\ze_R\big(s-1,{1\over2}\big)
-12\ze_R\big(s,{1\over2}\big)\bigg]+2^{1+s}.\cr}
$$
Combining the derivative at $s=0$ with (\peq{resi3}) according to
(\peq{qzetadash0}) gives
$$
W^D_{{1\over4}-3\rm
sphere}=-{1\over2}\ze'_{QS}(0)={1\over4}\ze'_R(-2)-{1\over24}\ze'_R(-1)-
{4\over3}\ln2-{1\over16}\approx-0.987416
$$
needed in ({\peq{3hemis}).

The calculation can be performed for any dimension $d$ and also for the
case when there are $q$ hyperplanes inclined at $\pi/q$.
\section{\bf 12. Conclusion}
Several extensions of this analysis can be envisaged. Higher dimensional
bundles, such as spinors and vectors, could be considered. This would be
connected with the recent work on mixed boundary conditions [\pref{Vass,
MandP,Kam2}].

A question we have not been able to resolve is the evaluation of the $\la$,
$\mu$ and $\nu$ functions in the expression for the heat-kernel
coefficient $C_{3/2}$ for general dihedral angles. There is continuing
interest in the calculation of such coefficients and it seems important
to the present authors to extend the analysis to the piecewise smooth
case, \cf [\pref{Cheeger}].

It should also be possible to
calculate the functional determinants on balls directly using properties
of Bessel functions, \cf [\pref{Barv}]. The heat-kernel coefficients on
balls have been computed by Stewartson and Waechter [\pref{SandW}] in
one dimension, Waechter [\pref{Waechter}] in two, and Kennedy
[\pref{Kennedy3}] up to five. We also mention the more recent work of
Moss [\pref{Moss}] and Bordag and Kirsten [\pref{BandK}]. The work of Berry
and Howls, [\pref{BandH}], extends that of Stewartson and Waechter to obtain
many more terms.

A technical problem of some value is the evaluation of the heat-kernel
coefficients and functional determinants on particular manifolds for the
general Robin case. Some results along these lines have been obtained by
Moss [\pref{Moss}] but could be extended. We note here that normalisable
zero modes might exist for specific values of $\psi$ which would affect the
calculation of the effective action.

\vskip 10truept
\noin{\bf{References}}
\vskip 5truept
\begin{putreferences}

\ref{KCD}{G.Kennedy, R.Critchley and J.S.Dowker \aop{125}{80}{346}.}
\ref{Donnelly}{H.Donnelly \ma{224}{1976}161.}
\ref{Fur2}{D.V.Fursaev {\sl Spectral geometry and one-loop divergences on
manifolds with conical singularities}, JINR preprint DSF-13/94,
hep-th/9405143.}
\ref{HandE}{S.W.Hawking and G.F.R.Ellis {\sl The large scale structure of
space-time} Cambridge University Press, 1973.}
\ref{DandK}{J.S.Dowker and G.Kennedy \jpa{11}{78}{895}.}
\ref{ChandD}{Peter Chang and J.S.Dowker \np{395}{93}{407}.}
\ref{FandM}{D.V.Fursaev and G.Miele \pr{D49}{94}{987}.}
\ref{Dowkerccs}{J.S.Dowker \cqg{4}{87}{L157}.}
\ref{BandH}{J.Br\"uning and E.Heintze \dmj{51}{84}{959}.}
\ref{Cheeger}{J.Cheeger \jdg{18}{83}{575}.}
\ref{SandW}{K.Stewartson and R.T.Waechter \pcps{69}{71}{353}.}
\ref{CandW}{H.S.Carslaw and J.C.Jaeger {\it The conduction of heat
in solids}
Oxford, The Clarendon Press 1959.}
\ref{BandH}{H.P.Baltes and E.M.Hilf {\it Spectra of finite systems}.}
\ref{Epstein}{P.Epstein \ma{56}{1903}{615}.}
\ref{Kennedy1}{G.Kennedy \pr{D23}{81}{2884}.}
\ref{Kennedy2}{G.Kennedy PhD thesis Manchester (1978).}
\ref{Kennedy3}{G.Kennedy \jpa{11}{78}{L173}.}
\ref{Luscher}{M.L\"uscher, K.Symanzik and P.Weiss \np {173}{80}{365}.}
\ref{Polyakov}{A.M.Polyakov \pl {103}{81}{207}.}
\ref{Bukhb}{L.Bukhbinder, V.P.Gusynin and P.I.Fomin {\it Sov. J. Nucl.
 Phys.} {\bf 44} (1986) 534.}
\ref{Alvarez}{O.Alvarez \np {216}{83}{125}.}
\ref{DandS}{J.S.Dowker and J.P.Schofield \jmp{31}{90}{808}.}
\ref{Dow1}{J.S.Dowker \cmp{162}{94}{633}.}
\ref{Dow2}{J.S.Dowker \cqg{11}{94}{557}.}
\ref{Dow3}{J.S.Dowker \jmp{35}{94}{4989}; erratum {\it ibid}, Feb.1995.}
\ref{Dow5}{J.S.Dowker {\it Heat-kernels and polytopes} To be published}
\ref{Dow6}{J.S.Dowker \pr{D50}{94}{6369}.}
\ref{Dow7}{J.S.Dowker \pr{D39}{89}{1235}.}
\ref{BandG}{P.B.Gilkey and T.P.Branson \tams{344}{94}{479}.}
\ref{Schofield}{J.P.Schofield Ph.D.Thesis, University of Manchester, 1991.}
\ref{Barnesa}{E.W.Barnes {\it Trans. Camb. Phil. Soc.} {\bf 19} (1903)
374.}
\ref{BandG2}{T.P.Branson and P.B.Gilkey {\it Comm. Partial Diff. Equations}
{\bf 15} (1990) 245.}
\ref{Pathria}{R.K.Pathria {\it Suppl.Nuovo Cim.} {\bf 4} (1966) 276.}
\ref{Baltes}{H.P.Baltes \prA{6}{72}{2252}.}
\ref{Spivak}{M.Spivak {\it Differential Geometry} vols III, IV, Publish
or Perish, Boston, 1975.}
\ref{Eisenhart}{L.P.Eisenhart {\it Differential Geometry}, Princeton
University Press, Princeton, 1926.}
\ref{Moss}{I.Moss \cqg{6}{89}{659}.}
\ref{Barv}{A.O.Barvinsky, Yu.A.Kamenshchik and I.P.Karmazin \aop {219}
{92}{201}.}
\ref{Kam}{Yu.A.Kamenshchik and I.V.Mishakov {\it Int. J. Mod. Phys.}
{\bf A7} (1992) 3265.}
\ref{Death}{P.D.D'eath and G.V.M.Esposito \prD{43}{91}{3234}.}
\ref{Rich}{K.Richardson \jfa{122}{94}{52}.}
\ref{Osgood}{B.Osgood, R.Phillips and P.Sarnak \jfa{80}{88}{148}.}
\ref{BCY}{T.P.Branson, S.-Y. A.Chang and P.C.Yang \cmp{149}{92}{241}.}
\ref{Vass}{D.V.Vassilevich.{\it Vector fields on a disk with mixed
boundary conditions} gr-qc /9404052.}
\ref{MandP}{I.Moss and S.Poletti \pl{B333}{94}{326}.}
\ref{Kam2}{G.Esposito, A.Y.Kamenshchik, I.V.Mishakov and G.Pollifrone
\prD{50}{94}{6329}.}
\ref{Aurell1}{E.Aurell and P.Salomonson \cmp{165}{94}{233}.}
\ref{Aurell2}{E.Aurell and P.Salomonson {\it Further results on functional
determinants of laplacians on simplicial complexes} hep-th/9405140.}
\ref{BandO}{T.P.Branson and B.\O rsted \pams{113}{91}{669}.}
\ref{Elizalde1}{E.Elizalde, \jmp{35}{94}{3308}.}
\ref{BandK}{M.Bordag and K.Kirsten {\it Heat-kernel coefficients of
the Laplace operator on the 3-dimensional ball} hep-th/9501064.}
\ref{Waechter}{R.T.Waechter \pcps{72}{72}{439}.}
\ref{GRV}{S.Guraswamy, S.G.Rajeev and P.Vitale {\it O(N) sigma-model as
a three dimensional conformal field theory}, Rochester preprint UR-1357.}
\ref{CandC}{A.Capelli and A.Costa \np {314}{89}{707}.}
\ref{IandZ}{C.Itzykson and J.-B.Zuber \np{275}{86}{580}.}
\ref{BandH}{M.V.Berry and C.J.Howls \prs {447}{94}{527}.}
\ref{DandW}{A.Dettki and A.Wipf \np{377}{92}{252}.}
\end{putreferences}
\newpage
\vskip 10truept\section{FIGURE CAPTIONS}
\noin Fig.1. \hspace{agai} Difference between the 2-cap and 2-hemisphere
effective actions
plotted against the colatitude of the cap rim, for Dirichlet boundary
conditions.
\vskip 7truept
\noin Fig.2. \hspace{agai} Difference between the 2-cap and 2-hemisphere
effective
actions
plotted against the colatitude of the cap rim, for Neumann boundary
conditions.
\vskip 7truept
\noin Fig.3. \hspace{agai} Difference between the 3-cap and 3-hemisphere
effective
actions
plotted against the colatitude of the cap rim, for Dirichlet boundary
conditions.
\vskip 7truept
\noin Fig.4. \hspace{agai} Difference between the 4-cap and 4-hemisphere
effective
actions
plotted against the colatitude of the cap rim, for Dirichlet boundary
conditions.
\vskip 7truept
\noin Fig.5. \hspace{agai} The effective action, W, on an equatorial
spherical
4-slice plotted against half its angular width, $\De$. (Dirichlet boundary
conditions.)
\vskip 7truept
\bye